\documentclass[journal,twoside,web]{ieeecolor}
\usepackage{generic}

\usepackage{mathrsfs}
\usepackage{color}
\usepackage{amsfonts}
\usepackage{subfigure}
\usepackage{booktabs}
\usepackage{cite}
\usepackage{amsmath,amssymb,amsfonts}
\usepackage{algorithmic}
\usepackage{graphicx}
\usepackage{pifont}
\usepackage{enumerate}

\usepackage[ruled]{algorithm2e} 
\usepackage{algorithmic}
\usepackage{bm}
\usepackage{makecell}
\usepackage{balance}
\usepackage{mathtools}
\usepackage{bbm}
\usepackage{dsfont}
\usepackage{pifont}

\usepackage{amsthm}

\usepackage{threeparttable}
\usepackage{textcomp}

\newtheorem{thm}{{Theorem}}
\newtheorem{lemma}{{Lemma}}
\newtheorem{cor}{{Corollary}}
\newtheorem{assumption}{{Assumption}}

\newtheorem{defn}{Definition}

\newtheorem{rem}{Remark}

\begin{document}
\title{Unidentifiability of System Dynamics: \\Conditions and Controller Design}
\author{Xiangyu Mao$^\dag$, \IEEEmembership{Student Member, IEEE} and Jianping He$^\dag$, \IEEEmembership{Senior Member, IEEE}
	\thanks{
	 $^\dag$: The Dept. of Automation, Shanghai Jiao Tong University, and Key Laboratory of System Control and Information Processing, Ministry of Education of China, Shanghai, China. E-mail address: \{maoxy20, jphe\}@sjtu.edu.cn. 
    Preliminary results have been published on the IFAC World Congress 2023 \cite{previ}.
	}
}
\maketitle

\begin{abstract}
How to make a dynamic system unidentifiable is an important but still open issue. It not only requires that the parameters of the systems  but also the equivalent systems cannot be identified by any identification approaches.
Thus, it is a much more challenging problem than the existing analysis of parameter identifiability. 
In this paper, we investigate the problem of dynamic unidentifiability and design the controller to make the system dynamics unidentifiable. 
Specifically, we first define dynamic unidentifiability by taking both system parameters and equivalent systems into consideration.
Then, we obtain the necessary and sufficient condition for unidentifiability based on the Fisher Information Matrix.
This condition is derived by analysis of the relationship between the unidentifiable parameters and the Hessian matrix of the system function.
Next, we propose a controller design scheme for ensuring dynamic unidentifiability under linear system models.
We prove that for controllable and observable linear time-invariant systems, the requirement of unidentifiability is equivalent to the requirement of a low-rank controller.
Then, the low-rank controller design problem is solved by transforming it into a model order reduction problem.
We demonstrate the effectiveness of our method by simulation.
\end{abstract}


\IEEEpeerreviewmaketitle

\section{Introduction}

With the continuous development and application of dynamic systems, information security has become a research hotspot.
Many of the attacks on a dynamic system, such as stealthy attacks, covert attacks, etc., rely on precise prior knowledge of the systems \cite{003,004,005,006,7866869}, which can be obtained by system identification.
Hence, we can enhance security significantly by designing a controller which prevents the adversary from identifying the dynamics of the systems \cite{desa}.
To achieve this objective, it is necessary to first derive the condition where the dynamics of the system are identifiable for the adversary.

Recently, identifiability for many system identification algorithms for both linear and nonlinear systems has been studied \cite{8362723,016}.
The condition for identifiability of system parameter values is given \cite{bellman_structural_1970}.
However, considering that we want to prevent the adversary from identifying the system dynamics, which means both the system parameters and equivalent systems (see Section III.A for detailed definition) cannot be identified by any identification approaches. 
Hence, it is a much more challenging issue than traditional identifiability analysis problems.

\subsection{Motivation}

Despite the prominent contributions of pioneering works on identifiability and controller design (see Section II for a review), there remain some notable issues in this field.
First, most of the existing research on system identification or system identifiability only gives sufficient conditions for system identifiability \cite{016, bellman_structural_1970, stru00,stru1}.
Few researches provide sufficient and necessary condition for the system to be unidentifiable from the perspective of information security.
For example, many of the studies of system identification methods use the condition that the input signal is a persistent excitation as the basic assumption that ensures the identifiability of the system \cite{003}.
Second, researches on identifiability mainly focus on parameterized systems and the parameter identifiability of systems \cite{012}. 
However, in practice, just knowing the dynamics of the system, i.e., the response of the system under a given exciting signal, is enough for the adversary to achieve a good performance of attack \cite{teixeira_data_2019,desa,phii}.
Hence, we need to consider all equivalent systems with different parameters but the same dynamics and prevent the adversary from identifying any of these equivalent systems.
Moreover, current works on optimal control with consideration of information security prefer using methods of adding noise to the controller to increase the identification error of the adversary \cite{domi}.
While the adversary may still be able to obtain convergent identification in noisy systems \cite{022}.
Hence, the condition for system unidentifiability and the controller design for unidentifiability remain open problems.

\subsection{Contribution}
To solve the above issues, we analyze the necessary and sufficient condition for the unidentifiability of system dynamics based on a parameterized system model.
Then, we design a controller to ensure the condition of unidentifiability to enhance the security of the system.
The main contributions are summarized as follows. 
 
\begin{itemize}
\item  For any parameterized system, we provide a necessary and sufficient condition for any parameter to be unidentifiable through the analysis of the Fisher Information Matrix (FIM).

\item  We obtain the necessary and sufficient condition for the dynamic identifiability of the system, which is different from the existing studies of parameter identifiability. 
We proved that the system is dynamically identifiable if and only if the null space of the FIM is a subset of the null space of the Hessian matrix.

\item Taking linear time-invariant (LTI) systems models as an example, we propose a controller design algorithm for LQR control while ensuring dynamic unidentifiability. We prove that for controllable and observable LTI systems, unidentifiable controller design problem for dynamic unidentifiability is equivalent to low-rank controller design problem.
Furthermore, we decompose the low-rank controller design problem into an LQR order reduction problem and propose an algorithm to solve it.

\end{itemize}
The theoretical results in this paper reveal the relationship between system dynamics and FIM of parameters and provide a controller design method to achieve unidentifiability.
Further research is called for exploring optimal controller design methods for general system models, such as nonlinear systems.

The differences between this paper and its conference version \cite{previ} include i) the identifiability of dynamics is investigated, ii) the extensions to system models under situations of other observation are given, iii) we provide a detailed analysis of the controller design problem which is based on the LTI system model and LQR minimization problem.

\subsection{Organization}
The remainder of this paper is organized as follows. 
Section \ref{Relatedwork} provides literature research on identifiability and controller design for unidentifiability.
We give the problem formulation, the assumptions, and notations in Sec. \ref{sec_preliminary}.
Then, the unidentifiability of system dynamics is investigated in Sec. \ref{sec_identifiability}, and the controller design algorithm is provided in Sec. \ref{sec_udesign}. 
Simulation results are shown in Sec. \ref{sec_sim}, followed by conclusions and future directions in Sec. \ref{sec_conclusion}.

\section{Related Work}\label{Relatedwork}
There has been extensive research on system identifiability and controller design for unidentifiability in the literature. 
This section gives a brief overview of them. 
 
\subsubsection{System identifiability}
In the literature, plenty of research effort has been devoted to the identifiability analysis of parameterized systems.
The concepts of identifiability are different according to different use scenarios, where the concepts often have overlapping definitions or equivalent definitions.
Some common concepts are practical identifiability \cite{On, Pra}, structural identifiability \cite{bellman_structural_1970,stru1,stru2}, and parameter identifiability.
In \cite{On}, practical identifiability is defined as the well-performed identification under the influence of noise, model uncertainty \cite{Pra}, or other disturbances. 
In \cite{bellman_structural_1970}, structural identifiability reflects the possibility of getting a system model from input-output measurement under a best-case scenario.
Structural identifiability requires that the observer is able to derive a unique solution or a finite, countable set of solutions of system parameters \cite{stru3,stru4,stru6}. 
Structural identifiability is considered an important property of systems and has been used in many researches \cite{stru7,stru8,stru9}, where the identifiability may have different names.
Moreover, structural identifiability can be explained in many different identification methods \cite{stru00}, such as the least-square method, the maximum likelihood method, etc.
In \cite{stru00}, these explanation versions of structural identifiability, such as least square identifiability, local identifiability via the information matrix, and identifiability in a consistency-in-probability sense, are investigated and the researcher finds the equivalence between them.
Hence, although system identifiability has various definitions, these definitions are usually common and equivalent.

Despite these definitions of identifiability, there are many researchers who focus on investigating the unidentifiability of systems.
For unidentifiable systems, an important research direction is to determine the identifiable part of the system.
In \cite{stru9}, the researchers determine the identifiable parameter combinations, and the functional forms for the dependencies between unidentifiable parameters via FIM.
In \cite{repa0,repa1}, a procedure for generating locally identifiable parts of unidentifiable systems is proposed by reparameterization.

Most of the existing research focuses on how to analyze systems that are already unidentifiable, while there are few researches on how to make systems unidentifiable. 
Therefore, this paper wants to analyze the necessary and sufficient conditions of unidentifiability and realize the unidentifiability.

\subsubsection{Controller Design}
To realize unidentifiability for the security of systems, controller design is an attractive way.
Recently, many different controller design methods have been proposed for achieving unidentifiability, where these methods consider different definitions of concepts of identifiability.
For example, \cite{myl} investigates the structural identifiability of linear time-invariant systems with an output feedback controller.
The researchers give the necessary and sufficient conditions where the adversary can identify the transfer function of the system.
Then, they design a low-rank controller against the Known-Plaintext Attack which renders the system unidentifiable to the adversary.
Another example is in \cite{ee},  the researchers study a controller design method for a linear unidentifiable single-input single-output system such that the attackers cannot identify the trajectory of the system.
These methods provide broad ideas for security based on unidentifiability.

Based on the existing works, this paper aims to achieve the unidentifiability of the system from multiple perspectives.
First, since the identifiability of each parameter and the quantitative definition of identifiability remain open issues, this paper gives the condition where each parameter is unidentifiable.
Second, a common issue in practice is that the adversary may not care about the exact value of parameters but care about the input-output dynamics of the system.
We define dynamic identifiability and investigate the condition of dynamic identifiability.
Then, we use controller design to make the adversary unable to identify the system dynamics.

\section{Preliminaries and Problem Formulation}\label{sec_preliminary}

This section gives the basic model investigated in this paper.
Then, we give the problem formulation and assumptions with some notations.

\subsection{Basic Model}
The basic model investigated in this paper is a parameterized discrete-time system, $\mathcal{S}(\bm{\theta}^*)$, which is given by 
\begin{equation}
    \mathcal{S}(\bm{\theta}^*) : \bm{y}(t) = \bm{f}(t,\bm{x},\bm{u},\bm{\theta}^*),
\end{equation}
where $t \in \mathbb{N}$, $\bm{y}(t) \in \mathbb{R}^{m}$ represents the output of $\mathcal{S}(\bm{\theta}^*)$,  $\bm{u} \in \mathbb{R}^{l\times (t+1)}$ is the input vector from time $0$ to $t$.
The term $\bm{x}$ is the internal parameter of $\mathcal{S}(\bm{\theta}^*)$, e.g., the state of $\mathcal{S}(\bm{\theta}^*)$ from time $0$ to $t$.
The parameter vector to be identified is $\bm{\theta}$ whose true value is $\bm{\theta}^*$.
The output function satisfies $\bm{f} = [f_1, f_2, \cdots, f_m]$, s.t., 
\[y_i(t) = f_i(t,\bm{x},\bm{u},\bm{\theta}^*),i = 1,2,\cdots,m,\]

We assume that the adversary identifies the parameters in a neighborhood of $\bm{\theta}^*$ in $\mathbb R^n$, i.e., the domain of the identification value of the parameters is $\mathcal{D}_{\bm{\theta}}$, which is a neighborhood of $\bm{\theta^*}$ in $\mathbb R^n$.
 We define the set of parameter values $\bm{\theta}$ that make $\mathcal{S}(\bm{\theta})$ have the same dynamics as $\mathcal{S}(\bm{\theta}^*)$ as $\Theta$, i.e.,
 \[
  \Theta = \{ \bm{\theta} | \bm{f}(t,\bm{x},\bm{u},\bm{\theta})  =  \bm{f}(t,\bm{x},\bm{u},\bm{\theta}^*),  \bm{\theta} \in \mathcal{D}_{\bm{\theta}} \}.
 \]
 Then, we give the definition of equivalent systems as follows.
 \begin{itemize}
    \item \textbf{Set of systems with the same dynamics (equivalent systems), $ \Psi$, defined by}
\end{itemize}
 \begin{equation}\label{eq_setofsamedynamic}
 \begin{aligned}
 \Psi = \{\mathcal{S}(\bm{\theta}) | \bm{\theta} \in \Theta\}.
\end{aligned}
 \end{equation}
Clearly, equivalent systems may have different system parameters while having the same output response under the same input signal.

\begin{table}[t]
\small
 \caption{\label{tab:test}Some Important Definitions}
 \begin{tabular}{ll}
 \toprule 
  Symbol  & Definition  \\ 
  \midrule
  $\mathcal{D}_{\bm{\theta}}$ &  domain of the identification value of the parameters;\\
  $W$ & the sensitivity matrix, see \eqref{eq_def_sensitivity};\\
  $F$ & the FIM, see \eqref{eq_def_fim};\\
  $J_a$ &  the Jacobian matrix, see \eqref{eq_def_jac};\\
  $H$ & the Hessian matrix, see \eqref{eq_def_Hessian};\\
  $\Psi$ & set of systems with the same dynamics, see \eqref{eq_setofsamedynamic};\\
  $\Psi_d$ & set of systems with the same dynamics \\
        & under given observation, see\eqref{eq_setofdynamicsdata};\\
  $\Psi_{J_a}$ & set of systems with the same Jacobian matrix, \\
         & see \eqref{eq_setofjaccobi};\\      
  $N(\cdot)$ & the null space of matrix  $(\cdot)$;\\
  $\|(\cdot)\|$ & the Frobenius of matrix $(\cdot)$;\\
  $\text{Span}(\cdot)$ & the linear space of vectors $(\cdot)$;\\
  \bottomrule 
 \end{tabular} \label{table-n0}
 \vspace{-5pt}
\end{table}

\subsection{Scenario and Definitions} 
Considering an infinite time process, the system has a trajectory/output sequence, $\bm{y}_d$, under a designed controller/control sequence $\bm{u}_d$, where $\bm{y}_d = [\bm{y}^\top(0), \bm{y}^\top(1), \cdots, \bm{y}^\top(t), \cdots]^\top$ and $\bm{u}_d = [\bm{u}^\top(0), \bm{u}^\top(1), \cdots, \bm{u}^\top(t),\cdots]^\top$.
Suppose that there is an adversary, who knows the system function form $\bm{f}$, the states $\bm{x}$, and the input and output data  $\{\bm{y}_d, \bm{u}_d\}$. 
The values of the parameters $\bm{\theta}^*$ are not available to the adversary.
We assume that the adversary passively observes the input and the output of the system.
The objective of the adversary is to use the observed data  $\{\bm{y}_d, \bm{u}_d\}$ to re-build the system model as $\hat{\mathcal{S}}$, where $\hat{\mathcal{S}}$ has the same dynamics as the original system $\mathcal{S}(\bm{\theta}^*)$.

Traditional system identification on parametric systems mainly focuses on identifying the exact value of system parameters\cite{stru4}.
Different from that, in this paper, the adversary only needs to know the dynamics of $\mathcal{S}(\bm{\theta}^*)$, i.e., derive $\Psi$.
For example, for a system described by a state-space model, the adversary only needs to identify the system up to a similarity transformation.

\begin{table}[t]
\small
 \caption{\label{tab:test}Relationships of the Definitions}
 \begin{tabular}{cccc}
 \toprule 
    & Parameters &  Dynamics & Relationships \\ 
  \midrule
  Identifiability & \ding{172} $\Psi_d = {S(\theta^*)}$ & \ding{173} $\Psi_d = \Psi$ & \ding{172}$\Rightarrow$\ding{173}\\
  Unidentifiability & \ding{174} ${S(\theta^*)} \subset \Psi_d$ & \ding{175} $\Psi \subset \Psi_d$ & \ding{174}$\Leftarrow$\ding{175}\\
  \bottomrule 
 \end{tabular} \label{table-n0}
\end{table}

\begin{figure}[t]
  \centering 
  \setlength{\abovecaptionskip}{0.1cm}
        \includegraphics[width=0.45\textwidth]{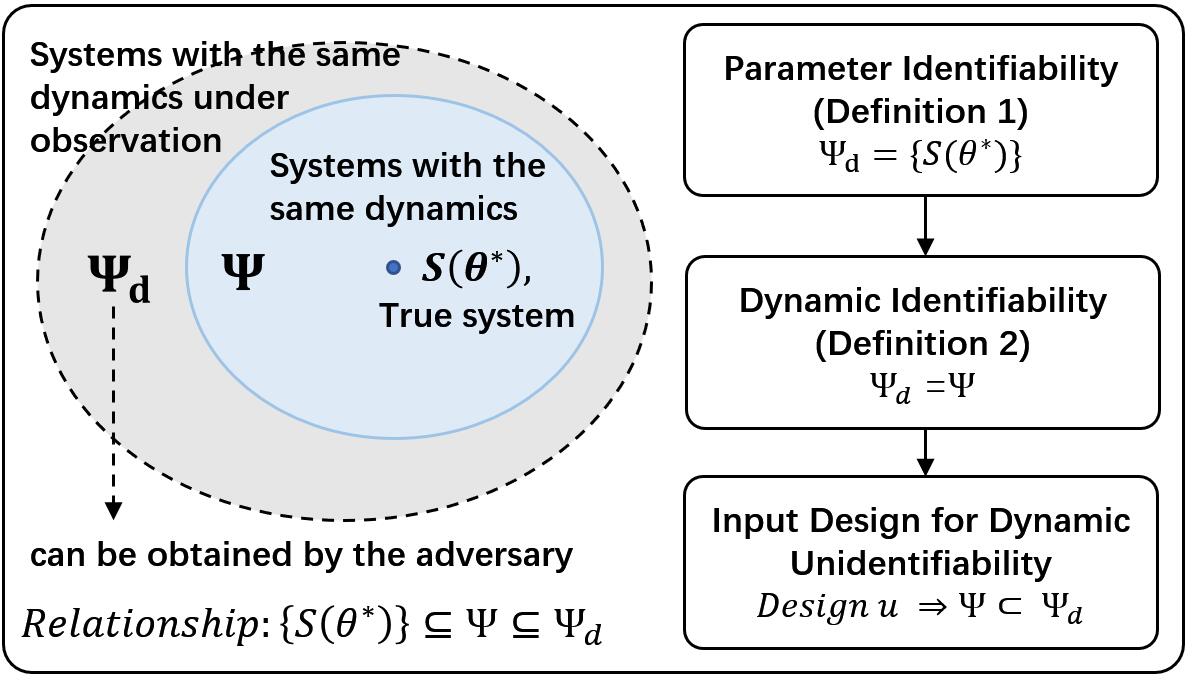} 
  \caption{Descriptions of definitions and problem formulation in this paper.} 
  \label{picPF}
  \vspace{-5pt}
\end{figure} 

This paper aims to design the control signal $\bm{u}_d$ to ensure that the adversary cannot derive $\Psi$.
We define the set of all the systems with the same dynamics as $\mathcal{S}(\bm{\theta}^*)$ under observed data as $\Psi_d$, i.e., 
 \begin{itemize}
    \item \textbf{Set of systems with the same dynamics under observation, $\Psi_d$, defined by} 
 \begin{equation}\label{eq_setofdynamicsdata}
 \begin{aligned}
      & \Psi_d = \{\mathcal{S}(\bm{\theta}) | \bm{\theta} \in \Theta_d \},      
 \end{aligned}
 \end{equation}
 where 
 \[\Theta_d = \{ \bm{\theta} | \bm{f}(t,\bm{x},\bm{u}_d,\bm{\theta})  =  \bm{f}(t,\bm{x},\bm{u}_d,\bm{\theta}^*) ,  \bm{\theta} \in \mathcal{D}_{\bm{\theta}} \}.\] 
\end{itemize}

It can be directly obtained from the definitions that $\Psi \subseteq \Psi_d$. 
We do not specify the method of the adversary for identification and assume the adversary uses the `best' identification method.
Thus, it can derive the exact $\Psi_d$ if there are no protect designs.

Then, we give the definitions of identifiability of system dynamics and parameters.

\begin{defn}[\textbf{Parameter identifiability}]
Given $\mathcal{S}, \bm{\theta}^*$, and $\bm{u}_d, \bm{y}_d$, the parameters $\bm{\theta}$ are (locally) identifiable iff $\exists \ \mathcal{D}_{\bm{\theta}}$,  s.t., $  \Psi_d  = \{\mathcal{S} (\bm{\theta}^*)\}$. 
\end{defn}

\begin{defn}[\textbf{Dynamic identifiability}]
Given $\mathcal{S}(\bm{\theta}^*)$, and $\bm{u}_d, \bm{y}_d$, the dynamics of system $\mathcal{S}(\bm{\theta}^*)$ are (locally) identifiable iff $\exists \ \mathcal{D}_{\bm{\theta}}$, s.t., $\Psi = \Psi_d$.
\end{defn}

\begin{rem}
The analysis of identifiability and unidentifiability is based on the given value of $\bm{\theta}$ at $\bm{\theta} = \bm{\theta}^*$ and is discussed under the condition that $\bm{\theta} \in \mathcal{D}_{\bm{\theta}}$, which is in line with the definition of local identifiability in literature\cite{stru00}.
If $\mathcal{D}_{\bm{\theta}} = \mathbb R^n$ in Definition 1. or Definition 2., the system is globally identifiable.
\end{rem}

We can directly give the definition of dynamic unidentifiability by Definition 2, which is the objective of this paper.
\begin{defn}[\textbf{Dynamic Unidentifiability}]
Given $\mathcal{S}, \bm{\theta}^*$,$ \mathcal{D}_{\bm{\theta}}$ and $\bm{u}_d, \bm{y}_d$, the dynamics of system $\mathcal{S}(\bm{\theta}^*)$ are (locally) unidentifiable iff $\Psi \subset \Psi_d$.
\end{defn}

\subsection{Problem Formulation}
First, we aim to analyze the condition for dynamic unidentifiability, i.e., find out when the system satisfies 
\[\Psi \subset  \Psi_d\].
Then, we compute the optimal input sequence $\bm{u}_d$ that optimizes a control problem while ensuring the dynamic unidentifiability in Definition 3.
We minimize a quadratic objective function $J(\bm{u}_d)$, which is given by
\begin{equation}\label{eq_init_u_design}
\begin{aligned}
P_1: \arg \min_{\bm{u}_d}  \ \  J(\bm{u}_d) = & \sum  \left( \bm{y}_d^\top(t)Q\bm{y}_d(t) + \bm{u}_d^\top(t)R\bm{u}_d(t)\right) ,\\
 s.t.,  \ \  & \Psi \subset  \Psi_d, \\
&  \bm{y}_d(t) = \bm{f}(t,\bm{x},\bm{u}_d,\bm{\theta}^*),
\end{aligned}
\end{equation}
where $Q, R \succeq \bm{0}$.

In Sec. \ref{sec_identifiability}, we focus on finding the necessary and sufficient condition of the constraint of unidentifiability, i.e.,  $\Psi \subset \Psi_d$.
Then, in Sec. \ref{sec_udesign}, we give the solution to \eqref{eq_init_u_design} to design the controller under an LTI system model.

The following assumption is made throughout the paper.
\begin{assumption}
The dynamic function $\bm{f}$ belongs to $C^1$ class (the class of first-order continuously differentiable functions) with respect to $\bm{\theta}$ and $\bm{u}$.
\end{assumption}
Assumption 1 is a basic guarantee for the feasibility of analysis of identifiability\cite{stru00}.

\begin{figure}[t]
  \centering 
  \setlength{\abovecaptionskip}{0.1cm}
    \includegraphics[width=0.48\textwidth]{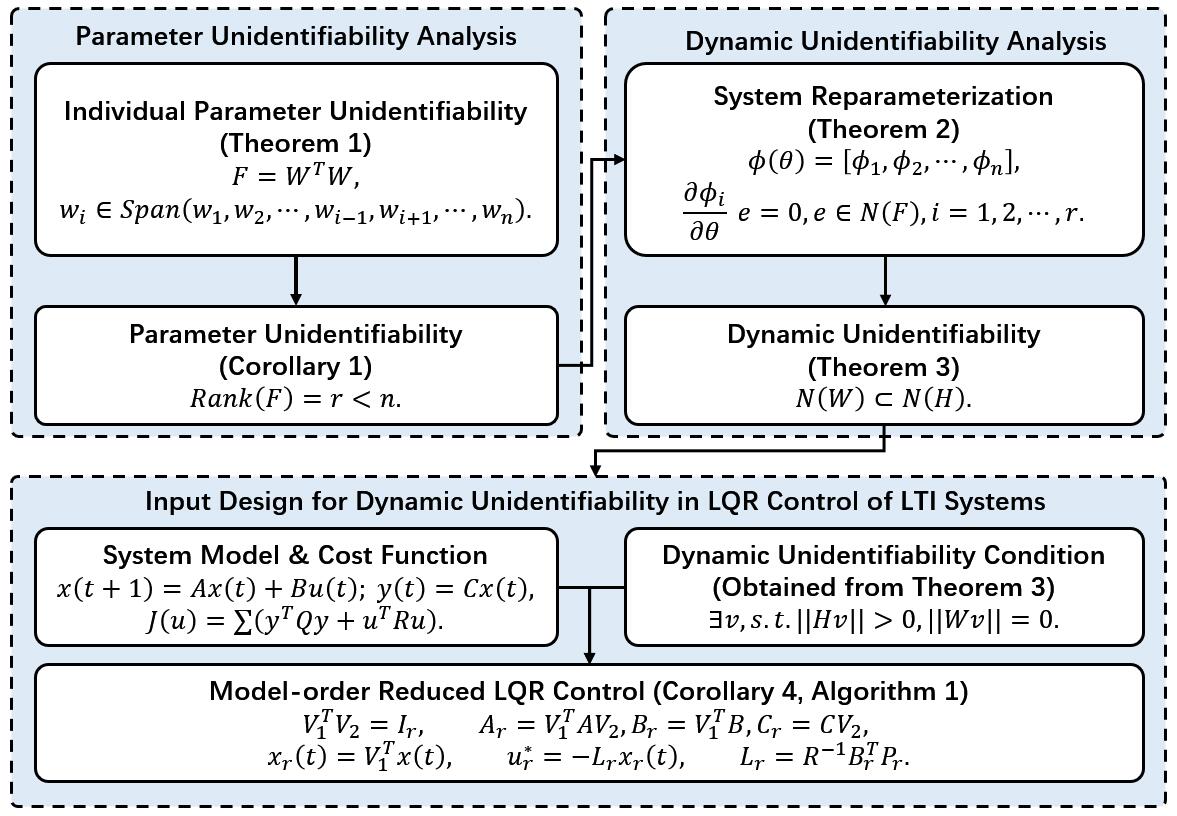} 
  \caption{Roadmap of the main theoretical results in this paper.} 
  \label{picroadmap}
  \vspace{-5pt}
\end{figure}

\section{System Unidentifiability}\label{sec_identifiability}

In this section, we investigate the condition of unidentifiability of the system dynamics.
We first start from the parameter unidentifiability via FIM.
Then, we analyze the relationship between parameter unidentifiability and dynamic unidentifiability, and give the necessary and sufficient condition of dynamic unidentifiability.

From the definitions of identifiability, we can easily derive the following lemma.
\begin{lemma}
Unidentifiability of parameters is a necessary condition for unidentifiability of the dynamics.
\end{lemma}
Hence, to investigate the condition of dynamic unidentifiability, we can start from investigating parameter unidentifiability.
Then, dynamic unidentifiability can be completed by deriving and avoiding the cases where parameters are unidentifiable but dynamics are identifiable.

\subsection{On the Unidentifiability of Parameters}

First, considering that $\mathcal{S}(\bm{\theta}^*)$ has $n$ parameters, we give the definition of the identifiability of each parameter.
\begin{defn}[\textbf{Individual parameter identifiability}]
Given $\mathcal{S}(\bm{\theta}^*)$ and $\bm{u}_d, \bm{y}_d$, the $i$-th parameter of  $\mathcal{S}(\bm{\theta}^*)$, $\theta_i$, is (locally) identifiable iff $\forall \bm{\theta} \in \Theta_d$, $\theta_i = \theta_i^*$. 
\end{defn}
\noindent It follows that the parameter identifiability in Definition 1 is equivalent to the identifiability of every individual parameter.

The FIM is a matrix that represents the amount of information contained in the observed data \cite{stru9}.
Hence, we use FIM to analyze the identifiability of parameters.
Given system $\mathcal{S}(\bm{\theta}^*)$ and observed data $\{\bm{y}_d, \bm{u}_d\}$, FIM is given by
\begin{equation}\label{eq_def_fim}
    F =  W^\top(\bm{y}_d, \bm{u}_d,\bm{\theta},\bm{\theta}^*) W(\bm{y}_d, \bm{u}_d,\bm{\theta},\bm{\theta}^*),
\end{equation}
where $W$ is the sensitivity matrix, a matrix with $n$ columns defined as
\begin{equation}\label{eq_def_sensitivity}
    W(\bm{y}_d, \bm{u}_d,\bm{\theta},\bm{\theta}^*\!) \!=\! 
    \begin{small}           
    \left[
\begin{array}{cccc}
     \frac{\partial \bm{y}_d(0)}{\partial \theta_1} & \frac{\partial \bm{y}_d(0)}{\partial \theta_2} & \cdots & \frac{\partial \bm{y}_d(0)}{\partial \theta_n}  \\
     \frac{\partial \bm{y}_d(1)}{\partial \theta_1} & \frac{\partial \bm{y}_d(1)}{\partial \theta_2} & \cdots & \frac{\partial \bm{y}_d(1)}{\partial \theta_n}  \\
    \vdots & \vdots & \vdots & \vdots \\
     \frac{\partial \bm{y}_d(t)}{\partial \theta_1} & \frac{\partial \bm{y}_d(t)}{\partial \theta_2} & \cdots & \frac{\partial \bm{y}_d(t)}{\partial \theta_n} \\
     \vdots & \vdots & \vdots & \vdots 
\end{array}\right] \end{small}
\Bigg |_{\bm{\theta} = \bm{\theta}^*}\!.
\end{equation}
We denote the $i$-th column of $W$ by $w_i$, then
\begin{equation}\nonumber
    W = [w_1, w_2, \cdots, w_n]. 
\end{equation}
Then, we have the following theorem.
\begin{thm}[\textbf{Unidentifiability of an individual parameter}]
\label{thm_para_uniden}
Provided $\text{rank}(F)$ keeps unchanged for all $\bm{\theta}$ in a neighbourhood of $\bm{\theta}^*$, an individual parameter $\theta_i$ is unidentifiable at $\bm{\theta}=\bm{\theta}^*$ iff
\[ w_i \in \text{Span}(w_1, \cdots,w_{i-1}, w_{i+1},\cdots, w_n).\]
\end{thm}
  
Theorem 1 gives a necessary and sufficient condition where each parameter $\theta_i$ is unidentifiable.
The proof of Theorem \ref{thm_para_uniden} is provided in Appendix A.
For the parameter unidentifiability of $\mathcal{S}(\bm{\theta}^*)$, we can directly obtain the following corollary.
\begin{cor}[\textbf{Unidentifiability of all parameters}]
\label{cor_para_uniden}
$\mathcal{S}(\bm{\theta}^*)$ is parameter unidentifiable iff there exist $\mathcal{D}_{\bm{\theta}}$ and $r < n$,  s.t.,  $\forall \bm{\theta} \in \mathcal{D}_{\bm{\theta}}$,  $\text{rank}(F) \equiv r$.
\end{cor}

 Corollary \ref{cor_para_uniden} provides the necessary and sufficient condition where $\mathcal{S}(\bm{\theta}^*)$ is unidentifiable in the perspective of parameters, i.e., the FIM is not full column rank, which is also necessary for the identifiability of system dynamics.

\subsection{On the Unidentifiability of System Dynamics}

First, for easier analysis, we give an another criterion for the dynamic identifiability.

We define the Jacobian matrix of the system $\mathcal{S}(\bm{\theta}^*)$ w.r.t $\bm{u}$ as $J_{a}(\bm{\theta})$, i.e.,
\begin{equation}\label{eq_def_jac}
    J_{a}(\bm{\theta}) = \left[
\begin{array}{cccc}
     \frac{\partial \bm{y}(0)}{\partial u_1} & \frac{\partial \bm{y}(0)}{\partial u_2} & \cdots & \frac{\partial \bm{y}(0)}{\partial u_n}  \\
     \frac{\partial \bm{y}(1)}{\partial u_1} & \frac{\partial \bm{y}(1)}{\partial u_2} & \cdots & \frac{\partial \bm{y}(1)}{\partial u_n}  \\
    \vdots & \vdots & \vdots & \vdots \\
     \frac{\partial \bm{y}(t)}{\partial u_1} & \frac{\partial \bm{y}(t)}{\partial u_2} & \cdots & \frac{\partial \bm{y}(t)}{\partial u_n} \\
     \vdots & \vdots & \vdots & \vdots \\
\end{array}\right],
\end{equation}
where $\frac{\partial \bm{y}(t)}{\partial u_i}$ represents the column vector of the derivative of $\bm{y}(t)$ for all $u_i(\tau)$ from $\tau = 0$ to $\tau = t$.
We define
 \begin{itemize}
    \item \textbf{Set of systems with the same Jacobian matrix, $\Psi_{J_{a}}$}
\end{itemize}
\begin{equation}\label{eq_setofjaccobi}
\begin{aligned}
    & \Theta_{J_{a}} = \{ \bm{\theta} |  J_{a}(\bm{\theta}) = J_{a}(\bm{\theta}^*),  \bm{\theta} \in \mathcal{D}_{\bm{\theta}} \}, \\
    & \Psi_{J_{a}} = \{\mathcal{S}(\bm{\theta}) | \bm{\theta} \in \Theta_{J_{a}} \}. 
\end{aligned}
\end{equation}
Then, we have the following lemma.
\begin{lemma}\label{lem_dynamic}
Given $\mathcal{S}(\bm{\theta}^*)$ and $\bm{u}_d, \bm{y}_d$, the dynamics of system $\mathcal{S}(\bm{\theta}^*)$ are (locally) identifiable iff $\exists \ \mathcal{D}_{\bm{\theta}}, \ s.t. \ \Psi_{J_{a}} = \Psi_d$.    
\end{lemma}
Please refer to Appendix B for the proof of Lemma \ref{lem_dynamic}.

Then, we investigate the gap between parameter unidentifiability and dynamic unidentifiability and find out the cases where the parameters are unidentifiable but the dynamics are identifiable.

From Lemma \ref{lem_dynamic}, we know that the dynamic unidentifiability is equivalent to the unidentifiability of the Jacobian matrix.
For a parameter-unidentifiable system, i.e., $\text{rank}(F) \equiv r$, previous studies have shown that locally identifiable parts can be decomposed from the unidentifiable parameters. 
Hence, we believe that when the Jacobian matrix can be described by the locally identifiable part, the dynamics are identifiable (which is explained in detail in our following theorems).
We give the following theorem for system reparameterization first.
\begin{thm}[\textbf{System reparameterization}]
\label{thm_repara}
Given $\mathcal{S}(\bm{\theta}^*)$ and $\bm{u}_d, \bm{y}_d$, provided there exist $\mathcal{D}_{\bm{\theta}}$ and $r < n$,  s.t.,  $\forall \bm{\theta} \in \mathcal{D}_{\bm{\theta}}$,  $\text{rank}(F) \equiv r$, then for all reparameterization function $\bm{\phi}(\bm{\theta}) = [{\phi}_1,{\phi}_2, \cdots,{\phi}_n ]^\top $, s.t.,
\begin{flalign}\nonumber
\begin{split}
1)  \ &  [\frac{\partial{\phi}_i}{\partial \theta_1},\frac{\partial{\phi}_i}{\partial \theta_2},\cdots, \frac{\partial{\phi}_i}{\partial \theta_n}] \cdot e = 0,   \forall e \in N(F),  i = 1,2,\cdots,r;\\
2) \ & \text{rank}([\frac{\partial\bm{\phi}}{\partial \theta_1},\frac{\partial\bm{\phi}}{\partial \theta_2},\cdots, \frac{\partial\bm{\phi}}{\partial \theta_n}]) = n, \\
\end{split}
\end{flalign}
we have $\{\phi_1, \phi_2,\cdots,\phi_r\}$ are $r$ identifiable parameters and $\{\phi_{r+1}, \phi_{r+2},\cdots,\phi_n\}$ are $(n-r)$ unidentifiable parameters of system $\mathcal{S}(\bm{\phi}(\bm{\theta}^*))$.
Moreover, for all $\bm{\phi}(\bm{\theta})$ satisfying 2),  system  $\mathcal{S}(\bm{\phi}(\bm{\theta}^*))$ has at most $r$ identifiable parameters.
\end{thm}
Please refer to Appendix C for the proof of Theorem \ref{thm_repara}.
Theorem \ref{thm_repara} is an extension of Theorem 3 in \cite{stru00}.
It gives us a reparameterization method to find the locally identifiable parameters of system $\mathcal{S}(\bm{\theta}^*)$.
For a simple example of the reparameterization function in Theorem \ref{thm_repara}, we assume that there exist $\mathcal{D}_{\bm{\theta}}$ and $r < n$,  s.t.,  $\forall \bm{\theta} \in \mathcal{D}_{\bm{\theta}}$,  $\text{rank}(F) \equiv r$, which means there exist $P\in \mathbb R^{n \times n}$, s.t.,
\begin{equation}\label{eq_P_of_repara}
  \text{rank}(P) = n, \ FP[\bm{0}_r, \ I_{n-r}] = \bm{0}_n.
\end{equation}
Then, we have the following corollary.

\begin{cor}\label{cor_repara}
For $P \in \mathbb R^{n \times n}$ satisfying \eqref{eq_P_of_repara} and reparameterization function $\bm{\phi}(\bm{\theta}) = [{\phi}_1,{\phi}_2, \cdots,{\phi}_n ]^\top  = P^\top\bm{\theta}$, we have $\{\phi_1, \phi_2,\cdots,\phi_r\}$ are $r$ identifiable parameters and $\{\phi_{r+1}, \phi_{r+2},\cdots,\phi_n\}$ are $(n-r)$ unidentifiable parameters of system $\mathcal{S}(\bm{\phi}(\bm{\theta}^*))$.    
\end{cor}

Corollary \ref{cor_repara} is an example of Theorem \ref{thm_repara} where the reparameterization function is a linear function of the parameters.
By the reparameterization in Corollary \ref{cor_repara}, we can analyze the dynamic identifiability in the following lemma.

\begin{lemma}\label{lem_reparadynamic}
Given $\mathcal{S}(\bm{\theta}^*)$ and $\bm{u}_d, \bm{y}_d$, provided there exist $\mathcal{D}_{\bm{\theta}}$ and $r < n$,  s.t.,  $\forall \bm{\theta} \in \mathcal{D}_{\bm{\theta}}$,  $\text{rank}(F) \equiv r$, the dynamics of $\mathcal{S}(\bm{\theta}^*)$ are (locally) identifiable iff for all $\bm{\phi}(\bm{\theta})$ satisfying 1) and 2) in Theorem \ref{thm_repara}, for every element of $J_{a}(\bm{\theta}^*)$ denoted by $J_{a_{ij}}$,
\begin{equation}\nonumber
    \frac{\partial J_{{a}_{ij}}}{\partial \phi_{r+k}} = \bm{0}, \  k = 0,1,\cdots,n-r, 
\end{equation}
\end{lemma}
Please refer to Appendix D for the proof of Lemma \ref{lem_reparadynamic}.
Lemma \ref{lem_reparadynamic} is an intuitive conclusion that combines the Jacobian matrix with the locally identifiable part.
It implies that after the reparameterization process in Theorem \ref{thm_repara}, if all the elements in the Jacobian matrix of the system are independent of unidentifiable parameters, the system is identifiable.

Considering that verifying the elements of $J(\bm{\theta}^*)$ has significant time complexity, we need to find uncomplicated methods.
We define the Hessian matrix of the system $\mathcal{S}(\bm{\theta})$ as $H(\bm{\theta})$, i.e.,
\begin{equation}\label{eq_def_Hessian}
    H(\bm{\theta}) =  \left[
\begin{array}{cccc}
     \frac{\partial^2 \bm{y}(0)}{\partial u_1 \partial \theta_1 } & \frac{ \partial^2 \bm{y}(0)}{\partial u_1 \partial \theta_2} & \cdots & \frac{\partial^2 \bm{y}(0)}{\partial u_1 \partial \theta_n}  \\
     \frac{\partial^2 \bm{y}(0)}{\partial u_2 \partial \theta_1} & \frac{\partial^2 \bm{y}(0)}{\partial u_2 \partial \theta_2} & \cdots & \frac{ \partial^2 \bm{y}(0)}{\partial u_2 \partial \theta_n}  \\
    \vdots & \vdots & \vdots & \vdots \\
     \frac{\partial^2 \bm{y}(0)}{\partial u_l \partial \theta_1} & \frac{\partial^2 \bm{y}(0)}{\partial u_l \partial \theta_2} & \cdots & \frac{\partial^2 \bm{y}(0)}{\partial u_l \partial \theta_n} \\
     \vdots & \vdots & \vdots & \vdots \\
     \frac{\partial^2 \bm{y}(t)}{\partial u_1 \partial \theta_1} & \frac{\partial^2 \bm{y}(t)}{\partial u_1 \partial \theta_2} & \cdots & \frac{\partial^2 \bm{y}(t)}{\partial u_1 \partial \theta_n} \\
     \vdots & \vdots & \vdots & \vdots \\     
\end{array}\right],
\end{equation}
where $\frac{\partial^2 \bm{y}(t)}{\partial u_i \partial \theta_j}$ represents the column vector of the second partial derivative of $\bm{y}(t)$ w.r.t. $u_i$ and $\theta_j$ for all $u_i(\tau)$ from $\tau = 0$ to $\tau  = t$ and a fixed $\theta_j$.
Then, we derive the following theorem.
\begin{thm}[\textbf{Identifiability of dynamics}]
\label{thm_NJNH}
Given $\mathcal{S}(\bm{\theta}^*)$ and $\bm{u}_d, \bm{y}_d$,  provided $\text{rank}(F)$ keeps unchanged for all $\bm{\theta}$ in a neighbourhood of $\bm{\theta}^*$, the dynamics of system $\mathcal{S}(\bm{\theta}^*)$ are (locally) identifiable iff $\forall  \bm{u}$,
\begin{equation}
    N(W(\bm{y}_d, \bm{u}_d,\bm{\theta},\bm{\theta}^*)) \subseteq N(H(\bm{\theta}^*)).
\end{equation} 
\end{thm}

\begin{rem}
    The Hessian matrix $H$ in Theorem \ref{thm_NJNH} is a matrix related to all the feasible control signal $\bm{u}$, where $\{\bm{u}_d\}$ is a subset of it.
    If $\{\bm{u}_d\}$ traverses all the value ranges of $\bm{u}$, it follows that $N(W(\bm{y}_d, \bm{u}_d,\bm{\theta},\bm{\theta}^*)) \subseteq N(H(\bm{\theta}^*))$ always holds.
    This is in line with intuition, where an attacker can obtain system dynamics if all the input and output data are observable. 
\end{rem}

Please refer to Appendix E for the proof of Theorem \ref{thm_NJNH}.
Theorem \ref{thm_NJNH} provides us a necessary and sufficient condition that the system is dynamically identifiable, i.e., the null space of the sensitivity matrix $W(\bm{y}_d, \bm{u}_d,\bm{\theta},\bm{\theta}^*)$ is equivalent to the null space of the Hessian matrix $H(\bm{\theta}^*)$.
Hence, for the controller design problem in \eqref{eq_init_u_design}, we can replace the constraint $\Psi \subset  \Psi_d$ by $N(W(\bm{y}_d, \bm{u}_d,\bm{\theta},\bm{\theta}^*)) \not\subseteq N(H(\bm{\theta}^*))$, which means
 \begin{itemize}
    \item \textbf{Condition for dynamics unidentifiability}
\end{itemize}
\begin{equation}\label{eq_newconstraint}
    \exists \ \bm{v} \in \mathbb R^n, \ s.t., \ \|W\bm{v}\| = 0, \|H\bm{v}\| > 0.
\end{equation}

Equation \eqref{eq_newconstraint} provides us with a feasible way to check the dynamic unidentifiability and design the controller to make the dynamics unidentifiable.

\subsection{Extension to Other Observation Models}

This subsection gives an extension of Theorem \ref{thm_NJNH} for dynamic unidentifiability.
We consider the scenario where an adversary only has part of the observation of $\bm{y}$, or the scenario where the adversary wants to predict the dynamic of $\varphi(\bm{y})$ rather than $\bm{y}$.
For these scenarios, we re-build the model of the problem formulated in Sec.\ref{sec_preliminary} as follows.
The parameterized discrete-time system $\tilde{\mathcal{S}}$ is given by 
\begin{equation}
\begin{aligned}
        \tilde{\mathcal{S}}(\bm{\theta}^*) :  \ & \bm{x}(t) = \tilde{\bm{f}}(t,\bm{u},\bm{\theta}^*),\\
        & \bm{y}(t) = \tilde{\bm{g}}(t,\bm{x}, \bm{u},\bm{\theta}^*), \\
        & \bm{z}(t) = \tilde{\bm{\varphi}}(t,\bm{x},\bm{u},\bm{\theta}^*),
\end{aligned}
\end{equation}
where $\bm{x}$ is the unobservable state of $\tilde{\mathcal{S}}$.
$\bm{y}$ and $\bm{u}$ are observable output and input data of $\tilde{\mathcal{S}}$.
The definitions of $\bm{\theta}$, $t$ and the domain of the identification value, $\mathcal{D}_{\bm{\theta}}$, are the same as the definitions  in Sec.\ref{sec_preliminary}.
$\bm{z}$ is the dynamics of system $\tilde{\mathcal{S}}(\bm{\theta}^*)$ which the adversary is supposed to identify, i.e.,  the set of systems that have the same dynamics with $\tilde{\mathcal{S}}(\bm{\theta}^*)$ is defined as
 \begin{equation}
     \tilde\Psi = \{ \tilde{\mathcal{S}}(\bm{\theta}) | \tilde{\bm{\varphi}}(t,\bm{x},\bm{u},\bm{\theta})  \equiv  \tilde{\bm{\varphi}}(t,\bm{x},\bm{u},\bm{\theta}^*),  \bm{\theta} \in \mathcal{D}_{\bm{\theta}} \}.
 \end{equation}

Then, we define the general sensitivity matrix $\tilde W$ and the general Hessian matrix $\tilde H$, where the matrix $\tilde W$ is defined the same as \eqref{eq_def_sensitivity} and $\tilde H$ is given by replacing $\bm{y}$ with $\bm{z}$ in \eqref{eq_def_Hessian}.

Next, we give the following theorem.
\begin{thm}\label{thm_extendedNJNH}
Given $\tilde{\mathcal{S}}(\bm{\theta}^*)$ and $\bm{u}_d, \bm{y}_d$,  provided $\text{rank}(\tilde W)$ keeps unchanged for all $\bm{\theta}$ in a neighbourhood of $\bm{\theta}^*$, the dynamics of system $\tilde{\mathcal{S}}(\bm{\theta}^*)$ are (locally) identifiable iff $\forall  \bm{u}$,
\begin{equation}
    N(\tilde W(\bm{y}_d, \bm{u}_d,\bm{\theta},\bm{\theta}^*)) \subseteq N(\tilde H(\bm{\theta}^*)).
\end{equation} 
\end{thm}
Hence, for general observation models, we have a necessary and sufficient condition of identifiability.

\section{Controller Design for Dynamic Unidentifiability of LTI Systems}\label{sec_udesign}

In this section, we use LTI systems as an example to illustrate the application of dynamic unidentifiability for security,
We propose a controller design algorithm for LTI systems to achieve optimal control while ensuring the system dynamics are unidentifiable.

The system model investigated in this section is a noise-free LTI model as follows.
\begin{equation}\label{eq_LTI_model}
\mathcal{S}(\bm{\theta}^*):
\begin{array}{ll}
\left\{\begin{array}{l}
\bm{x}(t+1) = A \bm{x}(t)+ B \bm{u}(t) \\
\bm{y}(t)= C  \bm{x}(t) ,
\end{array}\right.
\end{array}    
\end{equation}
where $x(t)\in\mathbb{R}^p$, $y(t)\in\mathbb{R}^m$, $u(t)\in\mathbb{R}^l$. 
We assume that the system is controllable and observable, and is with $0$ initial condition, i.e., $\bm{x}(0) = 0$.
System matrices are defined as $A \in\mathbb{R}^{p \times p}, B \in\mathbb{R}^{p \times l}$, $C \in   \mathbb{R}^{m \times p}$.
All the elements of $A, B, C$ are parameters to be identified and are part of $\bm{\theta}^*$.

\subsection{Analysis of LQR Control Problem with Unidentifiability}
\subsubsection{Problem formulation}
First, by Theorem \ref{thm_NJNH} and equation \eqref{eq_newconstraint}, we re-formulated the controller design problem $P_1$ as follows.
\begin{equation}\label{eq_lqr_u_design}
\begin{aligned}
P_1^1:~ \min_{\bm{u}_d}  \ \ & J(\bm{u}_d),\\
 s.t.,  \ \  & \exists \ \bm{v} \in \mathbb R^n,    \|W\bm{v}\| = 0, \|H\bm{v}\| > 0, \\
&  \bm{y}(t) = \bm{f}(t,\bm{x},\bm{u}_d,\bm{\theta}^*).
\end{aligned}
\end{equation}

We define the feasible set of $\bm{u}_d$ as $\mathcal{U}$, i.e., 
\begin{equation}\label{eq_feasible_u}
\begin{aligned}
    \mathcal{U} =  \{ \bm{u} | & \exists \ \bm{v} \in \mathbb R^n,  \|W\bm{v}\| = 0, \|H\bm{v}\| > 0,\\
&  \bm{y}(t)= C \bm{x}(t) ,\\
&  \bm{x}(t+1) = A  \bm{x}(t)+ B\bm{u}_d(t) \}.
\end{aligned}
\end{equation}
Hence, the controller design problem $P_1$ can be written as
\begin{equation}\label{eq_LTI_lqr_u_design}
\begin{aligned}
P_1^2:~ \min_{\bm{u}_d}  \ \ & J(\bm{u}_d),\\
 s.t.,  \ \  & \bm{u}_d \in \mathcal{U},
\end{aligned}    
\end{equation}
where $P_1$, $P_1^1$ and $P_1^2$ are equivalent problems.

\subsubsection{Calculation of Hessian matrix and sensitivity matrix}
Then, we derive the feasible set $\mathcal{U}$ to deal with the constraint in \eqref{eq_LTI_lqr_u_design}.
We assume that $\bm{y}_d$ and $\bm{u}_d$ are sequences from $t = 0$ to $t = T$.
Then, for any $k = 1, 2, \cdots, T$, by  $x(0) = 0$, we have 
\begin{equation}\nonumber
    y(k) = \sum_{i = 0}^{k-1} CA^{k-i-1}Bu(i)
\end{equation}
Next, we derive the Hessian matrix $H$ and the sensitivity matrix $W$ of $\mathcal{S}(\bm{\theta}^*)$ and investigate their relationship.
Define
\begin{equation}\nonumber
     g_{i,j,k} = \frac{\partial (CA^{k-j-1}B)}{\partial \theta_i}.
\end{equation}
We have the each element of $W$ denoting by
\begin{equation}\nonumber
    \frac{\partial y(k)}{\partial \theta_i} = \sum_{j = 0}^{k-1} g_{i,j,k} u(j),
\end{equation}
and the the each element of $H$ denoting by
\begin{equation}\nonumber
    \frac{\partial^2 y(k)}{\partial \theta_i \partial u(j)} = g_{i,j,k}.
\end{equation}
Hence, the matrices $W$ and $H$ are given by
\begin{equation}\nonumber
W  \!=\! 
    \begin{small}           
    \left[
\begin{array}{ccc}
      0 & \cdots & 0\\
      \sum_{j = 0}^{0} g_{1,j,0} u(j)  & \cdots &  \sum_{j = 0}^{0} g_{n,j,0} u(j)  \\
    \vdots & \vdots & \vdots \\
      \sum_{j = 0}^{k-1} g_{1,j,k-1} u(j) & \cdots &  \sum_{j = 0}^{k-1} g_{n,j,k-1} u(j) \\
     \vdots & \vdots  & \vdots 
\end{array}\right] \end{small}
\Bigg |_{\bm{\theta} = \bm{\theta}^*}\!,
\end{equation}
and
\begin{equation}\nonumber
 H =  \left[
\begin{array}{ccc}
     g_{1,0,0} & \cdots & g_{n,0,0}\\
     g_{1,1,0} & \cdots & g_{n,1,0}\\
     \vdots & \vdots & \vdots \\    
     g_{1,l-1,0} & \cdots & g_{n,l-1,0}\\
     g_{1,0,1} & \cdots & g_{n,0,1}\\
     \vdots & \vdots & \vdots \\ 
     g_{1,j,k} & \cdots & g_{n,j,k}\\
     \vdots & \vdots & \vdots \\ 
\end{array}\right],  
\end{equation}

\subsubsection{Low-rank controller for unidentifiability}
Next, we analyze the relationships between $W$ and $H$ and give the following theorem.
Define the set of low-rank controllers as $\mathcal{U}_K$, where 
\begin{equation}\label{eq_lowrank_set}
    \mathcal{U}_K = \{ \bm{u} |  \exists \ r < l,  K \in \mathbb R^{l \times r}, v \in  \mathbb R^{r}, \ \text{s.t.,} \  u = Kv \}.
\end{equation}
We have the following theorem.
\begin{thm}[\textbf{Low-rank controller for dynamic unidentifiability}]
\label{thm_lowrank_u}
For any LTI system model described in \eqref{eq_LTI_model}, we have
\begin{equation}\nonumber
   \mathcal{U}_K \subseteq \mathcal{U}.
\end{equation}
Furthermore, if the LTI system is controllable and observable, and the domain of $\bm{u}_d$ is a neighborhood, we have $\mathcal{U}_K = \mathcal{U}$.
\end{thm}
Please refer to Appendix F for the proof of Theorem \ref{thm_lowrank_u}.
Theorem \ref{thm_lowrank_u} means that a low-rank controller is a feasible solution for the unidentifiability of LTI systems.
Provided the rank of the controller, $r$, the controller design problem can be written as follows.
\begin{equation}\nonumber
\begin{aligned}
P_1^3:~ \  \min_{\bm{u}_d}  \ \ & J(\bm{u}_d),\\
 s.t.,  \ \  & \bm{u}_d  = Kv, \  v \in  \mathbb R^{r}\\
 & K \in \mathbb R^{l \times r}, \ r < l, \\
\end{aligned}        
\end{equation}
where $P_1^3$ is equivalent to $P_1$, $P_1^1$ and $P_1^2$ under our model.

\subsection{Low-rank Controller Design Algorithm}
Then, we solve the low-rank controller design problem $P_1^3$.

We derive the optimal $r$ first.
It can be easily inferred that for any $\{r_1, r_2\}$ satisfying $r_2 < r_1 < l$,
given $K_1 \in \mathbb R^{l \times r_1}$, $K_2 \in \mathbb R^{l \times r_2}$, $ v_1 \in  \mathbb R^{r_1}$,  $v_2 \in  \mathbb R^{r_2}$, we have
\begin{equation}\nonumber
    \min_{K_1, v_1} J(K_1 v_1) \leq \min_{K_2, v_2} J(K_2 v_2).
\end{equation}
Hence, the optimal $r$ is given by
\begin{equation}\nonumber
    r = l - 1.
\end{equation}

Next, we derive the optimal controller by substituting $\bm{u}_d = Kv$ in the minimization function $J$.

\subsubsection{Case 1: $p < l$}
First, we consider the case where the dimension of $x$ is smaller than $u$, i.e.,
$p < l$.
The solution of $\bm{u}_d$ to the LQR minimization problem \eqref{eq_lqr_u_design} is given by
\begin{equation}\label{eq_optimal_u_infinite}
 \bm{u}_d = - L_0 \bm{x}(t),
\end{equation}
where 
\begin{equation}\nonumber
    L_0 = ( R + B^\top P B)^{-1} B^\top P A,
\end{equation}
and $P \succeq \bm{0}$ is the solution to the following Riccati equation.
\begin{equation}\nonumber
\begin{aligned}
    A^\top P A \!-\! A^\top P B (R + B^\top P B ) ^{-1} B^\top P A  \!-\! P \!+\! C^\top Q C = 0.
\end{aligned}
\end{equation}
Notes that if $\bm{u} = - L_0 \bm{x}(t)$ and $p < l$, we can take $K = -L_0$ and $v = \bm{x}(t)$ in \eqref{eq_lowrank_set}, which means the optimal solution to the LQR problem is also a low-rank controller which makes the system dynamics unidentifiable.
Hence, we have the following corollary.
\begin{cor}\label{cor_lowranklqr}
For any LTI system with the model described in \eqref{eq_LTI_model} satisfying $p < l$,
denoting $\bm{u}^* = \arg \min_{\bm{u} \in \mathbb R^{l}} J(\bm{u})$ and $\bm{u}^*_d = \arg \min_{\bm{u} \in \mathcal{U}} J(\bm{u})$, we have $\bm{u}^* = \bm{u}^*_d$.
\end{cor}
\begin{rem}
If we consider the controller design problem within a finite time domain $[0, T]$, the cost function is given by
\begin{equation}\label{eq_costT_lqr}
    J_T(\bm{u}) \!=\! \sum_{t = 0}^{T-1} \left( \bm{y}^\top\!(t)Q\bm{y}(t) \!+\!\bm{u}^\top\!(t)R\bm{u}(t)\right) \!+\! \bm{y}^\top\!(T)Q_T\bm{y}(T),
\end{equation}
where $Q, R, Q_T \succeq \bm{0}$.                                                                                                                                                                                                                                                                                                
Then, we cannot derive a similar conclusion as Corollary \ref{cor_lowranklqr}.
Since in finite time domain $[0,T]$, the optimal solution to the LQR minimization problem is 
\begin{equation}\label{eq_optimal_u_finite}
 \bm{u}_d = - L_T(t) \bm{x}(t),
\end{equation}
where $L_T(t)$ is time variant.
\end{rem}

\begin{algorithm}[t]
		\small
 		\LinesNumbered
		\caption{Controller Design for Unidentifiability}\label{lowrankcontrolalgorithm}
		\For{$t \leqslant m$}
		{
                Simulate system $\mathcal{S}(\bm{\theta}^*)$ with random initial state $x(0)$.\\
                Record the system state in a finite time domain $[t_1, t_T]$.\\
                Build the snapshots matrix $X =[X, x_i(t_1), \cdots, x_i(t_T)]$.
	    }
           Compute the reduced SVD of $X = V_2 \Sigma V_1$.\\
           Obtain $x_r(t) = V_1^\top\bm{x}(t)$. \\
           Derive $P_r$ by the reduced Riccati equation \eqref{eq_reduced_riccati}.\\
           Obtain $L_r$ by $P_r$.\\
           Get the low-rank controller $\bm{u}_d$ by $\bm{u}_d = -  L_r x_r(t)$.
           
\end{algorithm}

\subsubsection{Case 2: $p \geq l$}
Second, we consider the case where $p \geq l$.
 Since $\bm{u} = Kv$ in $J(\bm{u})$, it follows that for any fixed $K$, we can derive the optimal $v = - L(K) \bm{x}(t)$, where
\begin{equation}\label{eq_KofLQR}
   L(K) = (K^\top R K  + K^\top B^\top P B K)^{-1}K^\top B^\top P A, 
\end{equation}
Then, the LQR minimization function can be formulated as an optimization function of $K$.
However, as we know that solving the Riccati equation is time-consuming and the optimization function of $K$ has no explicit form, the optimization problem is hard to handle directly.
Hence, we use a construction method based on model reduction to search $K$ for the minimization problem.

We assume that there exist basis matrices  $V_1, V_2 \in \mathbb R^{p \times r}$, where $ V_1^\top V_2 = I_r$, $ r < p$ and $r < l$.
It follows that if we let $\bm{x}(t) \approx V_2 \bm{x}_r(t)$,  the system model can be approximated by a low-rank model.
We define
\begin{equation}\nonumber
    A_r =  V_1^\top A V_2, \ B_r = V_1^\top B, \ C_r = C V_2, \ \bm{x}_r(t) = V_1^\top\bm{x}(t).
\end{equation}
Then, we have 
\begin{equation}\label{eq_LTI_reduced_model}
\mathcal{S}_r(\bm{\theta}^*):
\begin{array}{ll}
\left\{\begin{array}{l}
\bm{x}_r(t\!+\!1) = A_r \bm{x}_r(t)+ B_r \bm{u}(t) \\
\bm{y}_r(t)= C_r \bm{x}_r(t) ,
\end{array}\right.
\end{array}   
\end{equation}
It shows that we get a low-rank system model \eqref{eq_LTI_reduced_model} by the basis matrices $V_1, V_2$.
Denoting
\begin{equation}\label{eq_cost_reduced_lqr}
    J_r(\bm{u}) = \sum_{t = 0}^{\infty} \left( \bm{y}_r^\top(t)Q\bm{y}_r(t) + \bm{u}^\top(t)R\bm{u}(t)\right),
\end{equation}
where $\bm{y}$ satisfies the the order reduced model \eqref{eq_LTI_reduced_model}, we have the following corollary.
\begin{cor}\label{cor_modelreductionlqr}
For any LTI system $\mathcal{S}(\bm{\theta}^*)$ with order reduced model $\mathcal{S}_r(\bm{\theta}^*)$ given by \eqref{eq_LTI_reduced_model}, where $r<l $,
denoting $\bm{u}^*_r = \arg \min_{\bm{u} \in \mathbb R^{l}} J_r(\bm{u})$, we have $\bm{u}^*_r \in \mathcal{U}$.
\end{cor}

\begin{figure}[t]
  \centering 
  \setlength{\abovecaptionskip}{0.1cm}
    \includegraphics[width=0.45\textwidth]{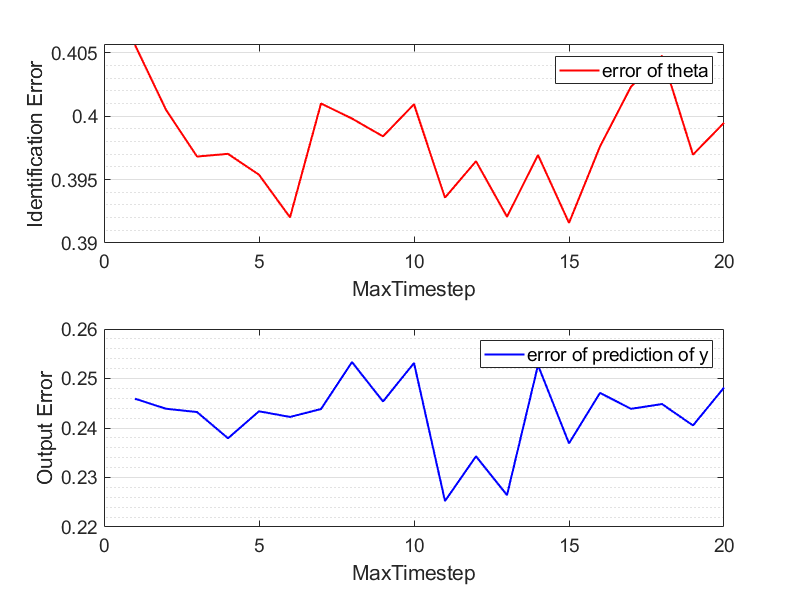} 
  \caption{The identification error of parameters and the prediction error of output of system $\mathcal{S}_1$ as the number of sample sets increases, obtained from 100 Monte Carlo runs with random noise.} 
  \label{picPara_i}
  \vspace{-5pt}
\end{figure}

\begin{proof}
    First, we derive the optimal controller in the LQR minimization problem of the order-reduced model.
    The controller $\bm{u}^*_r$ is given by
    \begin{equation}
        \bm{u}^*_r  = - L_r x_r(t),
    \end{equation}
    where $L_r$ is provided by 
    \begin{equation}\nonumber
        L_r = R^{-1}B_r^\top P_r,
    \end{equation}
    and $P_r\succeq \bm{0}$ is the solution to the following Riccati equation.
    \begin{equation}\label{eq_reduced_riccati}
    \begin{aligned}
        A_r^\top P_r A_r - &  A_r^\top P_r B_r (R + B_r^\top P_r B_r ) ^{-1} B_r^\top P_r A_r \\ & - P_r + C_r^\top Q C_r = 0.
    \end{aligned}
    \end{equation}
    Then, if we take $K = - L_r$ and $v(t) = x_r(t)$, we have $u(t) = Kv(t)$ holds for all $t$.
    Hence, we have $\bm{u}^*_r \in \mathcal{U}_K$.
    By Theorem \ref{thm_lowrank_u}, Corollary \ref{cor_modelreductionlqr} is proved.
\end{proof}
Corollary \ref{cor_modelreductionlqr} shows that if we can find basis matrices $V_1, V_2$ to reduce the order of the system model in the LQR minimization problem, the optimal controller of the order reduced system model also satisfies the requirement of unidentifiability.

There have been many studies and methods proposed on how to reduce the order of LQR problems.
This paper adopts the Proper Orthogonal Decomposition method (POD) for order reduction \cite{alla_order_2018}.

Finally, we propose our controller design algorithm using POD, which is given in Algorithm 1.
Note that for a general system model which may not satisfy $p < l$ and may not be controllable or observable, Algorithm 1 cannot ensure an optimal controller of the original minimization problem \eqref{eq_lqr_u_design}.

\section{Numerical Simulation}\label{sec_sim}

\begin{figure}[t]
  \centering 
  \setlength{\abovecaptionskip}{0.1cm}
    \includegraphics[width=0.45\textwidth]{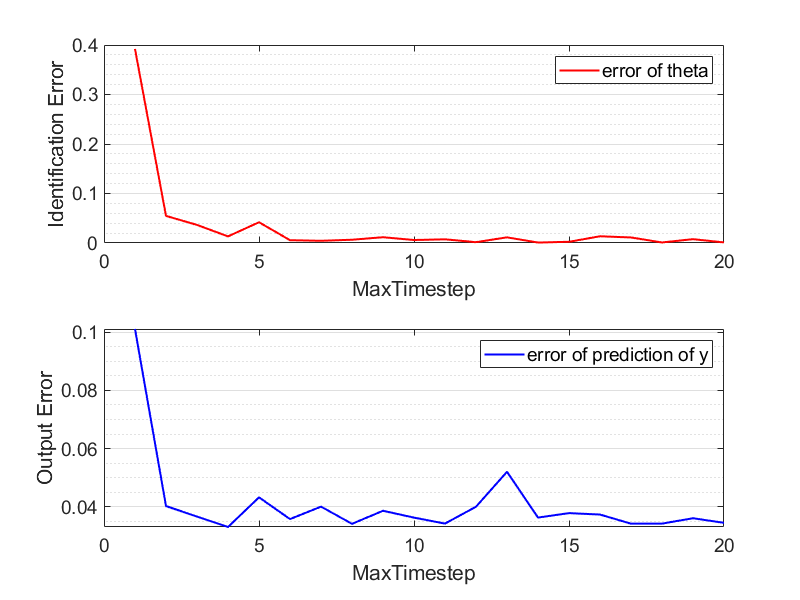} 
  \caption{The identification error of parameters and the prediction error of output of system $\mathcal{S}_2$ as the number of sample sets increases, obtained from 100 Monte Carlo runs with random noise.} 
  \label{picPara_u}
  \vspace{-5pt}
\end{figure}

This section uses numerical simulations to verify the effectiveness of the algorithms in this paper. 
The evaluation of unidentifiability is verified in the experiment in Figs. \ref{picPara_u} -  \ref{picDynamic}.
The control algorithm is verified in the experiment in  Fig. \ref{piccontrol}.

\subsection{Simulation Setting}
We randomly generate two LTI systems $\mathcal{S}_1$ and $\mathcal{S}_2$ where both of them are state-space models with 4 inputs, 4 states, and 4 outputs as follows.
\begin{equation}\nonumber
\mathcal{S}_1(\bm{\theta}^*):
\begin{array}{ll}
\left\{\begin{array}{l}
\bm{x}(t+1) = A(\bm{\theta}^*) \bm{x}(t)+ B(\bm{\theta}^*)\bm{u}(t) + w(t) \\
\bm{y}(t)= C(\bm{\theta}^*) \bm{x}(t) + v(t),
\end{array}\right.
\end{array}    
\end{equation}
\begin{equation}\nonumber
\mathcal{S}_2(\bm{\theta}^*):
\begin{array}{ll}
\left\{\begin{array}{l}
\bm{x}(t+1) = A(\bm{\theta}^*\!,\!\alpha) \bm{x}(t)+ B(\alpha)\bm{u}(t) + w(t)\\
\bm{y}(t)= C(\alpha) \bm{x}(t) + v(t),
\end{array}\right.
\end{array}    
\end{equation}
where $\alpha$ is a constant vector.
The parameters of $\mathcal{S}_1$ to be identified are all the elements in the parameter matrices $A$, $B$, $C$ of $\mathcal{S}_1$.
The parameters of $\mathcal{S}_2$ to be identified are the first row in parameter matrix $A$ of $\mathcal{S}_2$.
Other elements in the parameter matrices of $\mathcal{S}_2$, $\alpha$ are constants.
All the parameters and constants are randomly generated.
The noise is generated as a white noise sequence obeying uniform distribution.
The initial states of the two systems are all zeros.

\subsection{Evaluation of Unidentifiable Systems}

In this subsection, we verify the evaluation of parameter unidentifiability and dynamic unidentifiability.
In Fig. \ref{picPara_u} and \ref{picPara_i}, we use the stochastic gradient descent algorithm to identify the parameters of $\mathcal{S}_1$ and $\mathcal{S}_2$ and predict the output of the two systems by the identification of parameters.
First, from the definitions of the two systems, if the systems are noise-free, i.e., $w = v = 0$, we know that under a random controller with full rank, $\mathcal{S}_1$ is parameter unidentifiable but dynamic identifiable, and that $\mathcal{S}_2$ is both parameter identifiable and dynamic identifiable in theory.
In practice, when the systems have noise, we try to identify these two systems and evaluate the identifiability of them.
For generality, we identify the two systems with different parameters and different noise in 00 Monte Carlo runs and record the average error of identification of parameters or prediction of output.
The identification error of parameters and the prediction error of output of system $\mathcal{S}_1$ and $\mathcal{S}_2$ as the number of sample sets increases are presented in Fig. \ref{picPara_i} and \ref{picPara_u}, respectively.
It can be seen that as the size of data increases, the identification error of parameters and the prediction error of output of $\mathcal{S}_2$ converge to $0$.
However, the errors of $\mathcal{S}_1$ cannot converge to $0$.
On the one hand, since the parameter matrices of the MIMO system $\mathcal{S}_1$ are complicated for the stochastic gradient descent algorithm to identify, both the identification error of parameters and the output error are large, although the dynamic of $\mathcal{S}_1$ is identifiable.
On the other hand, for the simple system $\mathcal{S}_2$ which has only 4 parameters and is parameter identifiable, both the identification error and the output error quickly converge to $0$ after the time step reaches $5$.

\begin{figure}[t]
  \centering 
  \setlength{\abovecaptionskip}{0.1cm}
    \includegraphics[width=0.45\textwidth]{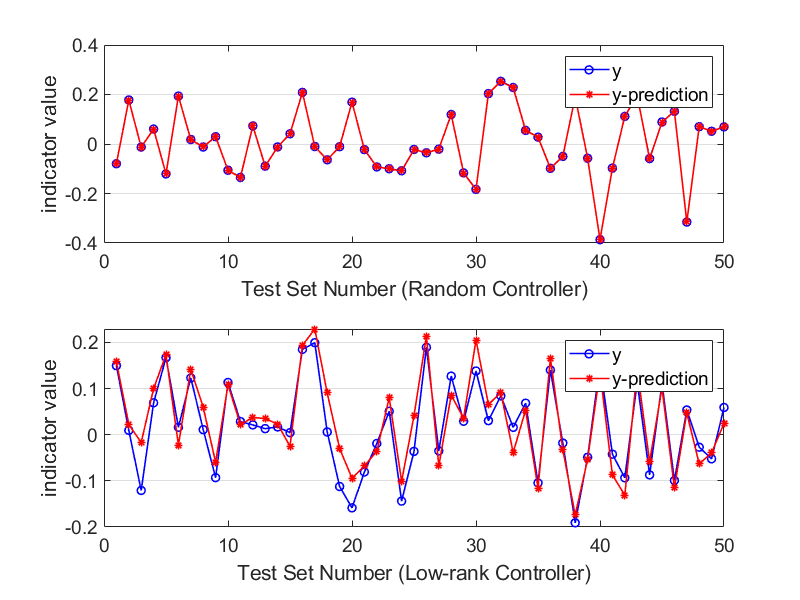} 
  \caption{The prediction error of output of system $\mathcal{S}_1$ under random controller and low-rank controller. The prediction is obtained by a back propagation neural network.} 
  \label{picDynamic}
  \vspace{-5pt}
\end{figure}

\subsection{Simulation of Controller Design Algorithm}

In Fig. \ref{picDynamic}, we use a backpropagation (BP) neural network to identify the dynamics of $\mathcal{S}_1$ under a random controller and a low-rank controller.
We use a three-layer fully connected neural network, with 15 nodes in the hidden layer for identification.
We randomly generated a control sequence with a maximum time step for $1000$ and generated the corresponding system output.
We use the first $950$ sets of input and output data as the training set, and the last $50$ sets of data as the testing set.
For the random controller, all the input sequences are randomly generated, which means the sequences are always with full rank.
For the low-rank controller, the input sequence of the training set is a rank-$3$ control signal, where we make the $4$-th dimension of input as a linear combination of other dimensions.
Meanwhile, the input sequence of the testing set is a full-rank control sequence.
It can be seen from Fig. \ref{picDynamic} that the BP neural network predicts the output of $\mathcal{S}_1$ precisely when we apply a random controller.
However, for a low-rank controller, the BP neural network cannot derive precise system dynamics.
Figure \ref{picDynamic} proves that the unidentifiability of the system provided by Theorem \ref{thm_lowrank_u} does not require a specific identification method, which demonstrates one of the advantages of our evaluation theorems.

\begin{figure}[t]
  \centering 
  \setlength{\abovecaptionskip}{0.1cm}
    \includegraphics[width=0.45\textwidth]{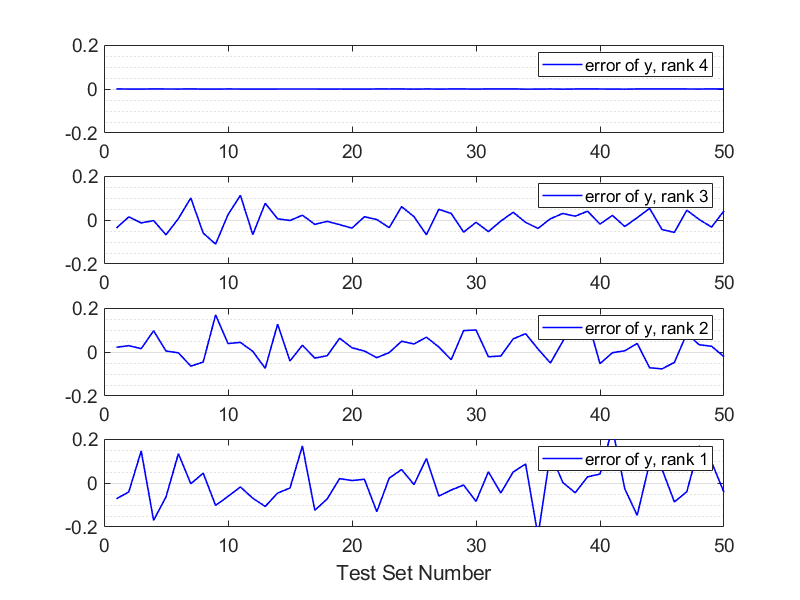} 
  \caption{The prediction error of output of system $\mathcal{S}_1$ under the controllers with different ranks. The prediction is obtained by a back propagation neural network.} 
  \label{piccontrol}
  \vspace{-5pt}
\end{figure}

In Fig. \ref{piccontrol}, we use BP neural network to identify the dynamics of $\mathcal{S}_1$ under controllers with different ranks to investigate the relationship between the rank of the controller and dynamic unidentifiability.
We adopt the same system model and network model as Fig. \ref{picDynamic}.
It shows that as the rank of the controller decreases, the identification error increases.
In practice, the low-rank controller affects the control performance of the system, so we need to make a trade-off between non Identifiability and control performance.

These simulation results demonstrate the effectiveness of the proposed algorithms.

\section{Conclusion}\label{sec_conclusion}

In this paper, we propose a controller design method for the unidentifiability of system dynamics.
We first study the identifiability of system dynamics.
We start with the investigation of the unidentifiability of parameters via the FIM.
It is found that we can reparameterize the system with several identifiable and unidentifiable parts.
Then, we analyze the relationship between the identifiability of system dynamics and parameters based on the reparameterization.
We finally come to the conclusion that the condition of dynamic identifiability is that, the null space of the sensitivity matrix is a subset of the null space of the Hessian matrix of the system.
Next, we design a controller that minimizes an LQR optimal control problem under an LTI system model while ensuring unidentifiability.
A low-rank controller is proved to be a feasible solution to the unidentifiability.
Then, the low-rank controller design problem is transformed into an LQR order reduction problem, which is solved by the Proper Orthogonal Decomposition method.
Finally, we reveal the effectiveness of the proposed algorithms by simulations.

Future directions include
i) analysis of the performance of the controller design algorithm;
ii) controller design for more general system models, such as nonlinear system models;
iii) considering system models with noise.

\section*{Acknowledgements}
The authors thank Yilin Mo and Na Li for insightful feedback and comments that improved the exposition of the paper.

\section*{Appendix}

\subsection{Proof of Theorem 1}

First, we prove the sufficiency, i.e., $\theta_i$ is unidentifiable if $w_i \in \text{Span}(w_1, \cdots,w_{i-1}, w_{i+1},\cdots, w_n)$.
We have
\begin{equation}\nonumber
    w_i = k_1 w_1 + \cdots + k_{i-1} w_{i-1}+ k_{i+1} w_{i+1} +\cdots+ k_n w_n,
\end{equation}
where $k_j$ are constants.  
Hence, for any $r \in \{1,2,\cdots,m\}$, we have
\begin{equation}\nonumber
   \frac{\partial f_r}{\partial \theta_i} = k_1 \frac{\partial f_r}{\partial \theta_1} + \cdots + k_{i-1} \frac{\partial f_r}{\partial \theta_{i-1}} + k_{i+1} \frac{\partial f_r}{\partial \theta_{i+1}}+\cdots+ k_n \frac{\partial f_r}{\partial \theta_n}.
\end{equation}
Then, since $\bm{\theta}^*$ is a regular point of $F$, we can always find $\Delta\theta_i$ and $\Delta\theta_j$, s.t.
\begin{equation}\nonumber
    f_r([\theta_1,\cdots,\theta_n]) = f_r([\theta_1 + \delta_1 \Delta\theta_1, \cdots, \theta_n + \delta_n \Delta\theta_n]), 
\end{equation}
where $\delta_j = 1, \forall j \neq i$ and $\delta_i = -1$.
This contradicts the description in Definition 2, which means $\theta_i$ is unidentifiable.

Next, we prove the necessity, i.e., if $\theta_i$ is unidentifiable, $w_i \in \text{Span}(w_1, \cdots,w_{i-1}, w_{i+1},\cdots, w_n)$.

By Definition 2, if $\theta_i$ is unidentifiable at $\bm{\theta}^*$, there exists $\bm{\theta}' \neq \bm{\theta}^* $ and $\theta_i' \neq \theta_i^*$ satisfying $\bm{f}(\bm{\theta}^*) = \bm{f}(\bm{\theta}')$.
By Assumption 2, $f_r$ and $\frac{\partial f_r}{\partial \theta_i}$ are continuous functions.
Then, according to the mean value theorem, there exists $\bm{\theta}^\dagger$ and $\Delta\theta_k$, s.t.
\begin{equation}\label{eq_proth1}
    \sum_{k = 1}^{n}\frac{\partial f_r(\bm{\theta}^\dagger)}{\partial \theta_k}\Delta\theta_k = 0.
\end{equation}
Since $\theta_i' \neq \theta_i^*$, $\Delta\theta_i \neq 0$.
Considering that $\bm{\theta}^*$ is a regular point of $F$, it follows that $\bm{\theta}^*$ also satisfies \eqref{eq_proth1}.
Hence, we have
\begin{equation}\nonumber
     \frac{\partial f_r}{\partial \theta_i} = \frac{1}{\Delta\theta_i} \sum_{j\neq i}\frac{\partial f_r}{\partial \theta_j}\Delta\theta_j.
\end{equation}
Therefore, Theorem 1 is proved.

\subsection{Proof of Lemma \ref{lem_dynamic}}

By the definition of dynamic identifiability in Definition 2, we have that the dynamics of $\mathcal{S}(\bm{\theta}^*)$ are (locally) identifiable iff $\exists \ \mathcal{D}_{\bm{\theta}}$, s.t., $\Psi = \Psi_d$.

Hence, to prove Lemma \ref{lem_dynamic} which says the dynamics of system $\mathcal{S}(\bm{\theta}^*)$ are (locally) identifiable iff $\exists \ \mathcal{D}_{\bm{\theta}}, \ s.t., \ \Psi_{J_{a}} = \Psi_d$, we just need to prove that 
\begin{equation}\nonumber
    \Psi = \Psi_{J_{a}},
\end{equation}
which means 
\begin{equation}\nonumber
    \mathcal{S}(\theta) \in \Psi \iff  \mathcal{S}(\theta) \in \Psi_{J_{a}}.
\end{equation}

First, we prove that
\begin{equation}\nonumber
    \mathcal{S}(\theta) \in \Psi \Rightarrow \mathcal{S}(\theta) \in \Psi_{J_{a}}.
\end{equation}
If $\mathcal{S}(\theta) \in \Psi$, we have that $\bm{f}(t,\bm{x},\bm{u},\bm{\theta})  =  \bm{f}(t,\bm{x},\bm{u},\bm{\theta}^*)$.
Hence, we have that for all $i = 1, 2, \cdots, l$,
\begin{equation}\nonumber
    \frac{\partial y}{\partial u_i}(\theta) =  \frac{\partial y}{\partial u_i}(\theta^*),
\end{equation}
which means $J_a(\theta) = J_a(\theta^*)$, thus, $\mathcal{S}(\theta) \in \Psi_{J_{a}}$.

Then, we prove that 
\begin{equation}\nonumber
    \mathcal{S}(\theta) \in \Psi_{J_{a}} \Rightarrow \mathcal{S}(\theta) \in \Psi.
\end{equation}
If $\mathcal{S}(\theta) \in \Psi_{J_{a}}$, we have that $ \frac{\partial y}{\partial u_i}(\theta) =  \frac{\partial y}{\partial u_i}(\theta^*)$.
By Assumption 1, we know that the system function $\bm{f}$ is first-order differentiable.
Hence, we have that,
\begin{equation}\nonumber
    \bm{f}(t,\bm{x},\bm{u},\bm{\theta})  =  \bm{f}(t,\bm{x},\bm{u},\bm{\theta}^*),
\end{equation}
which means $\mathcal{S}(\theta) \in \Psi$.

Therefore, Lemma \ref{lem_dynamic} is proved.

\subsection{Proof of Theorem \ref{thm_repara}}

First, given $\mathcal{S}(\bm{\theta}^*)$ with $\mathcal{D}_{\bm{\theta}}$ and rank of FIM, $r$, s.t., $r < n$,  and $\text{rank}(F) \equiv r$, provided a reparameterization function $\bm{\phi}(\bm{\theta}) = [{\phi}_1,{\phi}_2, \cdots,{\phi}_n ]^\top $, s.t.,
\begin{flalign}\nonumber
\begin{split}
  1)  \ &  [\frac{\partial{\phi}_i}{\partial \theta_1},\frac{\partial{\phi}_i}{\partial \theta_2},\cdots, \frac{\partial{\phi}_i}{\partial \theta_n}] \cdot e = 0,    \forall e \in N(F),   i = 1,\cdots,r \\
   2) \ & \text{rank}([\frac{\partial\bm{\phi}}{\partial \theta_1},\frac{\partial\bm{\phi}}{\partial \theta_2},\cdots, \frac{\partial\bm{\phi}}{\partial \theta_n}]) = n \\
\end{split}&
\end{flalign}
we prove that $\{\phi_1, \phi_2,\cdots,\phi_r\}$ are $r$ identifiable parameters.
For any $i = 1, 2, \cdots, n$, we have
\begin{equation}\nonumber
    \frac{\partial \bm{y}}{\partial \theta_i} =  \frac{\partial \bm{y}}{\partial \phi_1} \frac{\partial \phi_1}{\partial \theta_i}+ \frac{\partial \bm{y}}{\partial \phi_2} \frac{\partial \phi_2}{\partial \theta_i} + \cdots + \frac{\partial \bm{y}}{\partial \phi_n} \frac{\partial \phi_n}{\partial \theta_i}.
\end{equation}
It follows that
\begin{equation}\nonumber
\frac{\partial \bm{y}}{\partial \bm{\theta}}^\top = \frac{\partial \bm{y}}{\partial \bm{\phi}}^\top [\frac{\partial\bm{\phi}}{\partial \theta_1},\frac{\partial\bm{\phi}}{\partial \theta_2},\cdots, \frac{\partial\bm{\phi}}{\partial \theta_n}].
\end{equation}
Since $\text{rank}([\frac{\partial\bm{\phi}}{\partial \theta_1},\frac{\partial\bm{\phi}}{\partial \theta_2},\cdots, \frac{\partial\bm{\phi}}{\partial \theta_n}]) = n$, we have
\begin{equation}\nonumber
\frac{\partial \bm{y}}{\partial \bm{\phi}}^\top = \frac{\partial \bm{y}}{\partial \bm{\theta}}^\top [\frac{\partial\bm{\phi}}{\partial \theta_1},\frac{\partial\bm{\phi}}{\partial \theta_2},\cdots, \frac{\partial\bm{\phi}}{\partial \theta_n}]^{-1}.
\end{equation}
Hence, the sensitivity matrix of the reparameterized system $\mathcal{S}(\bm{\phi}(\bm{\theta}^*))$, $W'$ is given by
\begin{equation}\nonumber
    W' = W(y_d,u_d,\bm{\theta},\bm{\theta}^*) [\frac{\partial\bm{\phi}}{\partial \theta_1},\frac{\partial\bm{\phi}}{\partial \theta_2},\cdots, \frac{\partial\bm{\phi}}{\partial \theta_n}]^{-1}.
\end{equation}
Since $[\frac{\partial{\phi}_i}{\partial \theta_1},\frac{\partial{\phi}_i}{\partial \theta_2},\cdots, \frac{\partial{\phi}_i}{\partial \theta_n}] \cdot e = 0,   \ \forall e \in N(F), \   i = 1, 2, \cdots, r$, we have that $(w'_1, w'_2, \cdots, w'_r)$ is a set of bases of linear space $\text{Span}(w_1, w_2, \cdots, w_n)$.
Hence, $\{\phi_1, \phi_2,\cdots,\phi_r\}$ are $r$ identifiable parameters of system $\mathcal{S}(\bm{\phi}(\bm{\theta}^*))$.

Then, since $\text{rank}([\frac{\partial\bm{\phi}}{\partial \theta_1},\frac{\partial\bm{\phi}}{\partial \theta_2},\cdots, \frac{\partial\bm{\phi}}{\partial \theta_n}]) = n$, we have
\begin{equation}\nonumber
  \text{rank}(W') = \text{rank}(W(y_d,u_d,\bm{\theta},\bm{\theta}^*)) = r.
\end{equation}
Combining the above proof that $(w'_1, w'_2, \cdots, w'_r)$ is a set of bases of linear space $\text{Span}(w_1, w_2, \cdots, w_n)$, we know that
\begin{equation}\nonumber
    w'_{r+i} \in \text{Span}(w'_1, w'_2, \cdots, w'_r), \ i = 1,2,\cdots,n-r.
\end{equation}
Hence, by Theorem 1, $(w'_{r\!+\!1}, \cdots, w'_n)$ are $(n-r)$ unidentifiable parameters of the system.
The proof of $\mathcal{S}(\bm{\phi}(\bm{\theta}^*))$ having at most $r$ identifiable parameters can be directly obtained from $ \text{rank}(W') = r$.

Hence, Theorem \ref{thm_repara} is proved.

\subsection{Proof of Lemma \ref{lem_reparadynamic}}

From Lemma \ref{lem_dynamic}, we know that the dynamics of $\mathcal{S}(\bm{\theta}^*)$ are identifiable iff $\Psi_{J_{a}} = \Psi_d$.

Providing $\bm{\phi}(\bm{\theta})$ satisfying (1)-(2) in Theorem \ref{thm_repara},  we consider constructing a system 
\begin{equation}\nonumber
    \mathcal{S}': \bm{y}' = \bm{f}'(t,\bm{x}', \bm{u}, \bm{\phi}(\bm{\theta}^*)),
\end{equation}
where
\begin{equation}\nonumber
    \bm{y}' = [J_{{a}_{11}}, J_{{a}_{12}}, \cdots, J_{{a}_{nn}}]^\top.
\end{equation}

Then, we have that the dynamics of $\mathcal{S}(\bm{\theta}^*)$ are identifiable iff  $\mathcal{S}'$ is parameter identifiable.
By Corollary \ref{cor_para_uniden},  $\mathcal{S}'$ is parameter identifiable iff the FIM of $\mathcal{S}'$, $F'$ has full rank.
By Theorem \ref{thm_repara}, we know that 
\begin{equation}\nonumber
    \text{rank}(F') = r.
\end{equation}

Since $\bm{\phi}(\bm{\theta}) = [{\phi}_1,{\phi}_2, \cdots,{\phi}_n ]^\top $, the parameters of system $\mathcal{S}'$ are (locally) identifiable iff 
\begin{equation}\nonumber
    \frac{\partial y'}{\partial \phi_{r+k}} = \bm{0}, \  k = 0,1,\cdots,n-r, 
\end{equation}

Hence, Lemma \ref{lem_reparadynamic} is proved.

\subsection{Proof of Theorem \ref{thm_NJNH}}

We use Lemma \ref{lem_reparadynamic} to prove Theorem \ref{thm_NJNH}.

First, we use the example of the reparameterization function in Sec. \ref{sec_identifiability}.
we assume that there exists $P\in \mathbb R^{n \times n}$, s.t.,
\begin{equation}\nonumber
  \text{rank}(P) = n, \ FP[\bm{0}_r, \ I_{n-r}] = \bm{0}_n.
\end{equation}
Then, we have the reparameterization function $\bm{\phi}(\bm{\theta}) = [{\phi}_1,{\phi}_2, \cdots,{\phi}_n ]^\top  = P^\top\bm{\theta}$.

By Lemma \ref{lem_reparadynamic}, the dynamics of system $\mathcal{S}(\bm{\theta}^*)$ are (locally) identifiable iff $ \frac{\partial J_{a_{ij}}}{\partial \phi_{r+k}} = \bm{0}, \  k = 0,1,\cdots,n-r$.
It follows that
\begin{equation}\nonumber
  \frac{\partial^2 y}{\partial u \partial \phi_{r+k} } = \bm{0}, \  k = 0,1,\cdots,n-r.
\end{equation}
Since
\begin{equation}\nonumber
\frac{\partial \bm{y}}{\partial \bm{\phi}}^\top = \frac{\partial \bm{y}}{\partial \bm{\theta}}^\top P^{-1},
\end{equation}
we have that $\mathcal{S}(\bm{\theta}^*)$ is dynamics identifiable iff
\begin{equation}\nonumber
     \frac{\partial^2 \bm{y}}{\partial \bm{u} \partial \bm{\theta} } = [ W_r,\bm{0}_{n-r}]P^{-1},
\end{equation}
where $W_r$ is a constant matrix determined by $\frac{\partial^2 y}{\partial u \partial \phi_{k} }, \  k = 1,\cdots,r$.
Hence, the dynamics of $\mathcal{S}(\bm{\theta}^*)$ are identifiable iff $\forall x \in \mathbb R^{n-r}$,
\begin{equation}\nonumber
    P[\bm{0}_r, x_{n-r}] \subseteq N(H(\bm{\theta}^*)).
\end{equation} 
From the fact that
\begin{equation}\nonumber
  \text{rank}(P) = n, \ FP[\bm{0}_r, \ I_{n-r}] = \bm{0}_n,
\end{equation}
we have $ N(W(\bm{y}_d, \bm{u}_d,\bm{\theta},\bm{\theta}^*)) = P[\bm{0}_r, x_{n-r}]$.
Hence, the dynamics of $\mathcal{S}(\bm{\theta}^*)$ are identifiable iff
\begin{equation}\nonumber
    N(W(\bm{y}_d, \bm{u}_d,\bm{\theta},\bm{\theta}^*)) \subseteq N(H(\bm{\theta}^*)).
\end{equation} 
Theorem \ref{thm_NJNH} is proved.

\subsection{Proof of Theorem \ref{thm_lowrank_u}}
First, we prove that for any LTI system $\mathcal{S}(\bm{\theta}^*)$ in \eqref{eq_LTI_model},
Since the each element of $W$ is given by
\begin{equation}\nonumber
    \frac{\partial y(k)}{\partial \theta_i} = \sum_{j = 0}^{k-1} g_{i,j,k} u(j),
\end{equation}
and the the each element of $H$ is given by
\begin{equation}\nonumber
    \frac{\partial^2 y(k)}{\partial \theta_i \partial u(j)} = g_{i,j,k},
\end{equation}
by Theorem \ref{thm_NJNH}, the dynamics of $\mathcal{S}(\bm{\theta}^*)$ are unidentifiable iff
\begin{equation}\nonumber
    \exists \ \bm{v} \in \mathbb R^n, \ s.t., \ \|W\bm{v}\| = 0, \|H\bm{v}\| > 0.
\end{equation}
Denoting $(\cdot)_k$ as the $k$-th element of vector $(\cdot)$, we have
\begin{equation}\nonumber
    (W\bm{v})_k = \sum_i \sum_j g_{i,j,k}u(j) v_i,
\end{equation}
and
\begin{equation}\nonumber
    (H \bm{v})_k =  \sum_i g_{i,j,k} v_i.
\end{equation}
Hence, the dynamics of $\mathcal{S}(\bm{\theta}^*)$ are unidentifiable iff
\begin{equation}\nonumber
    \exists \ \bm{v} \in \mathbb R^n, \ s.t., \ \sum_i \sum_j g_{i,j,k}u(j) v_i = 0, \sum_i g_{i,j,k} v_i \neq 0.
\end{equation}
If $\bm{u} \in \mathcal{U}_K$, which means $\bm{u} = K \bm{z}$, where $dim(\bm{z}) < dim(\bm{u})$.
Then, we can always find $\bm{v}$ satisfying $\sum_i \sum_j g_{i,j,k}K \bm{z} v_i = 0$.
Hence, the system is dynamic unidentifiable, which means $\mathcal{U}_K \subseteq \mathcal{U}$.

Then, if the LTI system is controllable and observable, if $\bm{u} \not\in \mathcal{U}_K$, we have
\begin{equation}\nonumber
    \text{rank}(\bm{u}_d(1), \cdots, \bm{u}_d(T)) = dim(\bm{u}).
\end{equation} 
Since $\mathcal{S}(\bm{\theta}^*)$ is controllable, and the domain of $\bm{u}_d$ is a neighborhood, there exists a trajectory $\bm{u}_d(1), \cdots, \bm{u}_d(T)$ satisfying
\begin{equation}\nonumber
    \text{rank}(\left[
\begin{array}{cc}
     \bm{u}_d(1), \cdots, \bm{u}_d(T)  \\
     \bm{x}(1), \  \cdots, \ \bm{x}(T)  \\
\end{array}\right]) =  dim(\bm{u}) +  dim(\bm{x}),
\end{equation} 
which means $\bm{u}_d$ is a persistently exciting.
Then, the dynamics of $\mathcal{S}(\bm{\theta}^*)$ are identifiable by traditional identification methods.
Hence, for controllable and observable LTI systems, if the domain of $\bm{u}_d$ is a neighborhood, we have $\mathcal{U}_K = \mathcal{U}$.

\balance
\bibliographystyle{IEEEtran}

\bibliography{ref} 

\begin{IEEEbiographynophoto}{Xiangyu Mao}
(S'21) received the B.E. degree in Department of Automation from Tsinghua University, Beijing, China, in 2020. 
He is currently working toward the Ph.D. degree with the Department of Automation, Shanghai Jiaotong University, Shanghai, China. 
He is a member of Intelligent of Wireless Networking and Cooperative Control group. 
His research interests include system identification, networked systems and  distributed  optimization in multi-agent networks. 
\end{IEEEbiographynophoto}

\begin{IEEEbiographynophoto}{Jianping He} (SM’19) is currently an associate professor in the Department of
Automation at Shanghai Jiao Tong University. He received the Ph.D. degree in control science and engineering from Zhejiang University, Hangzhou, China, in 2013, and had been a research fellow in the Department of Electrical and Computer Engineering at University of Victoria, Canada, from Dec. 2013 to Mar. 2017. His research interests mainly include the distributed learning,
control and optimization, security and privacy in network systems.

Dr. He serves as an Associate Editor for IEEE Trans. Control of Network Systems, IEEE Open Journal of Vehicular Technology, and KSII Trans. Internet and Information Systems. He was also a Guest Editor of IEEE TAC, IEEE TII, International Journal of Robust and Nonlinear Control, etc. He was the winner of Outstanding Thesis Award, Chinese Association of Automation, 2015. He received the best paper award from IEEE WCSP'17, the best conference paper
award from IEEE PESGM'17, and was a finalist for the best student paper award from IEEE ICCA'17, and the finalist best conference paper award from IEEE VTC'20-FALL.
\end{IEEEbiographynophoto}

\end{document}